# New magnetic moment theory may explain features of solar jets and spicules

Leonard Freeman
leonard.freeman@ntlworld.com

## Abstract

Current theory states that the magnetic moment of a charged particle is constant, or invariant in a slowly changing magnetic field. It also states that the magnetic flux through a Larmor orbit is constant.

The current theory is closely examined, and found to have inconsistencies. An alternative theory is developed with new results for both the energy and magnetic moment of a charged particle. The new theory predicts particle behaviour which is opposite to convention: for example, that a plasma will decelerate as it moves through a decreasing magnetic field. Conversely it will accelerate into an increasing field. The magnetic field of the sun would then act as a restraining influence on solar plasma ejections.

The new theory is developed for two forms of the magnetic field: an inverse square law and an exponential. The theoretical results are compared with recent experimental results for the velocity, deceleration, time and heights of solar jets and spicules. The effect on Coronal Mass Ejections and Coronal Bright Fronts is also examined.

## 1. Introduction

When a charged particle moves at right angles to a uniform magnetic field, it rotates in Larmor circles. If this magnetic field is steadily increased the current conventional theory claims that the magnetic flux enclosed by the orbit remains constant, and also that the kinetic energy of the particle increases in proportion to the field.

It is suggested here that these statements are incompatible, and that this is due to inconsistencies in the conventional theory, both in its derivation and conclusions.

It is well known that when a charged particle, moving at velocity V at right angles to a magnetic field B, then the radius, r, of the Larmor circle is given by

$$r = \frac{mV}{Be} \qquad (1)$$

where m=particle mass and e=particle charge

If the magnetic field slowly increases then the radius r will slowly decrease as shown in figure 1.

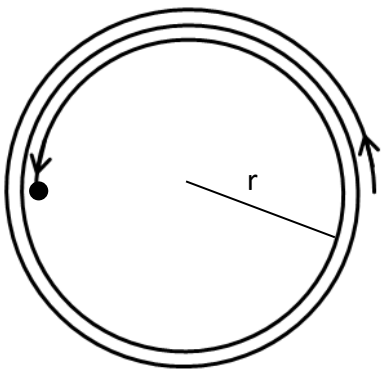

**Figure 1**

As the magnetic field increases the radius, r, decreases.

A changing magnetic field will also cause an induced emf around each orbit, due to electromagnetic induction, so that the energy and velocity, V, of the particle will also change.

## 2  Conventional theory

This was first suggested by Alfven (1950), and the proof is reproduced on the following page.
The main condition is that rate of change of B must be slow, that is the relative change in B during one single orbit must be much less than one, so that the orbits are almost circular.
The main results of conventional theory state that:

- **The magnetic flux, Φ, through the Larmor orbit is constant. ie**

$$\Phi = \text{constant} \tag{2}$$

- **The kinetic energy of the particle, $W_\perp$, is proportional to the applied field, B, ie**

$$W_\perp = \mu B \tag{3}$$

(Where μ is a constant, depending on the initial conditions and $W_\perp$ is the kinetic energy at right angles to the field, in the plane of the Larmor orbit.)

The change in energy is caused by the change in magnetic flux through the orbit of the particle.

### Current theory – an example

This theory can be outlined by considering the effect of a large change in magnetic flux density, B, from $B_0$ to $10 B_0$, for example. If B is changing slowly, then the particle will then execute a large number of revolutions during this time, of gradually decreasing radius.

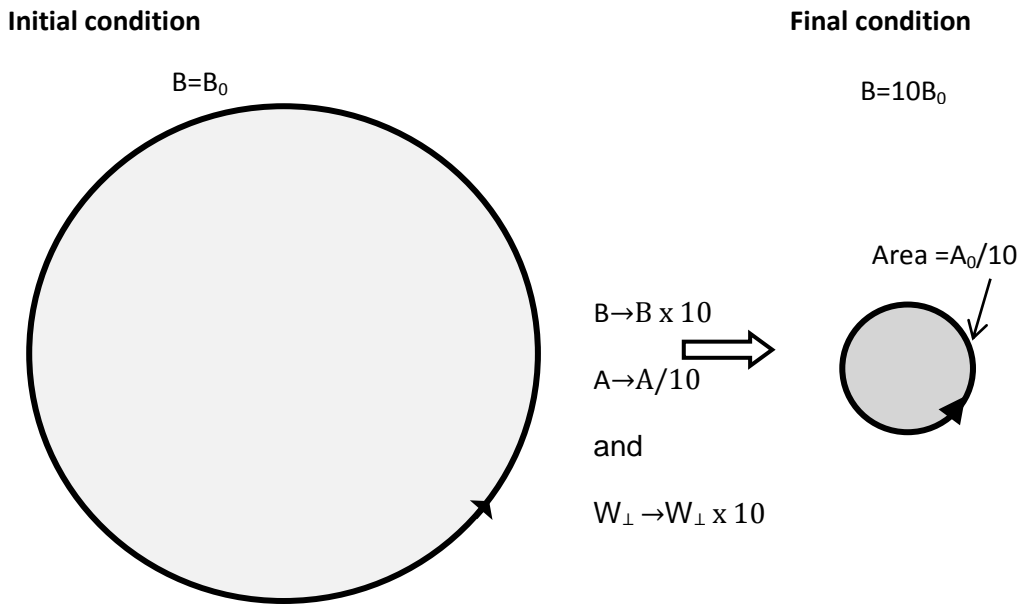

**Initial condition**            **Final condition**

B=$B_0$            B=$10B_0$

Area =$A_0/10$

B→B x 10

A→A/10

and

$W_\perp$ → $W_\perp$ x 10

For a uniform magnetic field, B, at right angles to the area, A,
the magnetic flux, Φ, is given by:  $\Phi = B A$  (4)
So  $\delta\Phi = B\delta A\delta + A\delta B$

**Now if equation (2) is correct then** $\delta\emptyset = 0$ **and so current theory requires that**

$$\frac{\delta A}{A} = \frac{-\delta B}{B} \qquad (5)$$

So Alfven's result means that, in order for the magnetic flux to remain constant, the relative change in area of a Larmor orbit is of equal importance as the change in magnetic flux density.

## Derivation of the conventional Alfven theory

Alfven's proof starts with the emf induced around the particle's orbit which, to quote:

*"changes the energy of the particle. We have*

$$\oint E.dl = -\frac{d\emptyset}{dt} \qquad (6)$$

$\emptyset \ (= \pi r^2 B)$ *is the flux through the circular path of the particle and the integral is to be taken along the periphery of the same circle."*

*"The gain in energy in one turn is*

$$\delta W_\perp = -e\oint E.dl = e\pi r^2 \frac{dB}{dt} \qquad (7)$$

*"(The negative sign derives from the fact that a positive particle goes in a direction opposite to that in which the integral is to be taken.) Thus the rate of increase in energy is given by":*

$$\frac{dW_\perp}{dt} = \frac{\delta W_\perp}{T} = \frac{W_\perp}{B}\frac{dB}{dt}$$

Alfven then uses equation (1), and the formula for the periodic time in a Larmor orbit,

$$T = \frac{2\pi}{\omega} = \frac{2\pi m}{Be} \qquad (8)$$

and by eliminating dt and integrating, gets the conventional result:

$$W_\perp = \mu B \qquad (9)$$

where μ, the magnetic moment, is a constant.
(Note that the magnetic moment, μ is defined as the ratio $W_\perp/B$)
It then follows that the magnetic flux through a gyro-orbit, Φ must also be constant.

This can easily be seen as follows:

$$\Phi = BA = B\pi r^2 = B\pi \frac{m^2 V_\perp^2}{B^2 e^2} = \pi \frac{m^2 V_\perp^2}{Be^2} \qquad (10)$$

which is clearly constant if $\frac{mV_\perp^2}{B} =$ constant, i.e. if the magnetic moment is constant.

## 2. Contradictions and errors in the current theory

It can be seen in equation (7) that Alfven gives $\frac{d\emptyset}{dt}$ as $\pi r^2 \frac{dB}{dt}$ ie as $A\frac{dB}{dt}$,

rather than $A\frac{dB}{dt} + B\frac{dA}{dt}$, where A is the area of the Larmor orbit.

**Two problems with current theory**

1. The change in area between two orbits has clearly been neglected. Alfven himself does not mention this assumption, although other authors who reproduce this proof do mention this point, for example Gartenhaus (1964). However, as shown earlier, current theory implies that the change in area is as important as the change in flux density.

2. In Alfven's equations (6) and (7), the energy gain δW over one Larmor orbit is calculated (usingan integral form of Faraday's induction law), by finding the rate of change in magnetic flux, $\emptyset/dt$ . But according to the **result** of conventional theory, (equation 2), the magnetic flux does not change – it is constant even for large changes in B. So on the one hand, the theory requires a small flux change between orbits, while on the other hand, the result of the theory is that there is no flux change! In fact equation (3) permits **very large** changes in the kinetic energy of the particle, but equation (2) states that this happens with **no change** in magnetic flux: so where does the induced electric field come from?

These contradictions in Alfven's theory arise because of the assumption that the change in area, from the end of one orbit to the end of the next, can be ignored. This area is shown in figure 2b:

Figure 2a
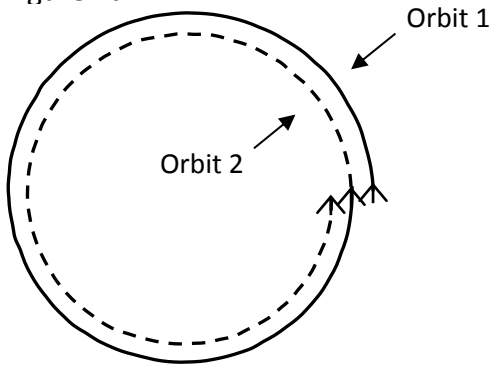
Orbit 1
Orbit 2

Figure 2b
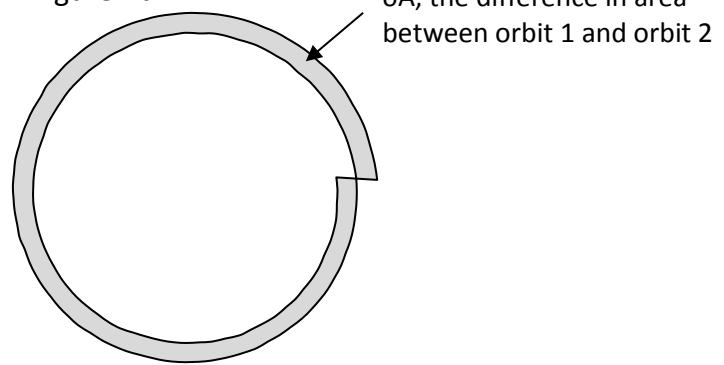
δA, the difference in area between orbit 1 and orbit 2

## 3. An alternative derivation and new theory

The following derivation attempts to take into account the effect not only of a change in magnetic flux, but also of the associated change in area of the Larmor orbit.

The main equation is the integral form of the Faraday/Maxwell Induction law:

$$\oint \mathbf{E} \cdot d\mathbf{l} = -\frac{d}{dt} \int \mathbf{B} \cdot d\mathbf{s} \qquad (11)$$

**B** and d**s** are the magnetic flux density and elemental area vectors, which in figure 9 both point away from the viewer. The right hand rule then means that the positive direction for **E** and d**l** are clockwise, as indicated.

The left hand side of equation (11) is the induced emf, ε, around the Larmor orbit. The right hand side is often expressed simply as $-d\emptyset/dt$. The effect of the minus sign is that if $d\emptyset/dt$ is positive, then the induced electric field and the emf will act in an anti-clockwise direction in the case shown in figure 9, where the direction of B is away from the viewer, and at right angles to the plane of the orbit.

Figure 9

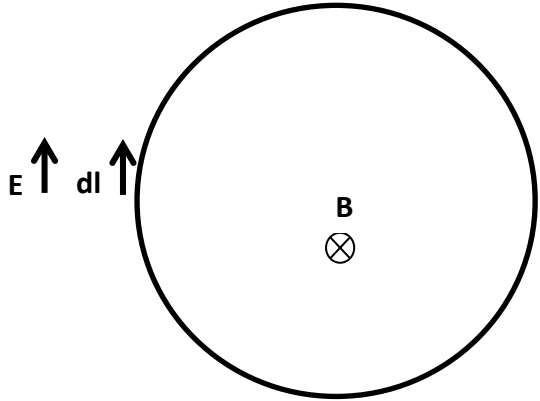

If the magnetic field, B is uniform over the area A of the Larmor orbit we simply get:

$$\int \mathbf{B} \cdot d\mathbf{s} = \mathbf{B} \cdot \mathbf{A} = BA$$

The flux through a Larmor orbit is given by $\Phi = BA$
Then the rate of change of flux is given by

$$\frac{d\Phi}{dt} = B\frac{dA}{dt} + A\frac{dB}{dt}$$

In conventional theory the rate of change of area is taken as zero, but it has been shown that this factor is as important as the change in field, B and so cannot be neglected.

For the moment, for simplicity we will consider a (hypothetical) positive particle that is rotating clockwise in figure 9, that is, one with a positive velocity vector in the same positive direction as both **E** and d**l**.
If the velocity of the particle is taken as V then the area A is found by substituting from equation (1):

$$A = \pi r^2 = \pi \frac{m^2 V^2}{B^2 e^2}$$

so $\quad \dfrac{dA}{dt} = 2\pi r \dfrac{dr}{dt}$

Taking B and V, as usual, to be time dependent, and substituting for r,

$$\frac{dA}{dt} = 2\pi \frac{m^2 V}{Be^2}\left(-\frac{V}{B^2}\frac{dB}{dt} + \frac{1}{B}\frac{dV}{dt}\right)$$

so

$$B\frac{dA}{dt} = \frac{\pi m^2 V}{Be^2}\left(-2\frac{V}{B}\frac{dB}{dt} + 2\frac{dV}{dt}\right)$$

And the emf is given by

$$\varepsilon = -\left(B\frac{dA}{dt} + A\frac{dB}{dt}\right) = \frac{\pi m^2 V}{Be^2}\left(\frac{V}{B}\frac{dB}{dt} - 2\frac{dV}{dt}\right) \tag{12}$$

The energy gain over one rotation is $e\varepsilon$, and using the period as given by equation (8), the rate of change of energy is

$$\frac{d}{dt}\left(\frac{mV^2}{2}\right) = \frac{e\varepsilon}{T} = \frac{\varepsilon Be^2}{2\pi m}$$

Using (12) and (8) to eliminate $\varepsilon$ and T gives:

$$\frac{d}{dt}\left(\frac{mV^2}{2}\right) = \frac{mV}{2}\left(\frac{V}{B}\frac{dB}{dt} - 2\frac{dV}{dt}\right)$$

Or

$$\frac{dW_\perp}{dt} = \frac{W_\perp}{B}\frac{dB}{dt} - \frac{dW_\perp}{dt}$$

So

$$\frac{dW_\perp}{dt} = \frac{W_\perp}{2B}\frac{dB}{dt} \tag{13}$$

However, this is for a hypothetical positive particle, rotating clockwise in figure 9. A real positive particle will rotate in the **opposite** direction, so will experience an electric field in the opposite direction. If the hypothetical particle is accelerating, then a real particle will be decelerating by the same amount, and vice-versa. So for a real particle we must have:

$$\frac{dW_\perp}{dt} = -\frac{W_\perp}{2B}\frac{dB}{dt} \tag{14}$$

Solving this gives:

$$W_\perp = \frac{k}{B^{\frac{1}{2}}} \tag{15}$$

where k is a constant, given by the initial conditions for W and B.

This means that an increase in the magnetic flux density, B, will lower the energy of the particle. This is completely opposite to the conventional theory, where an increase in B produces an increase in W.

This means that the magnetic moment, defined as $\mu \equiv \frac{W_\perp}{B}$, is not constant, as conventional theory requires. From (15) it can be seen that

$$\mu = \frac{k}{B^{3/2}} \qquad (16)$$

*Cautionary note: differential form of Maxwell/Faraday law*

The differential form, sometimes used in conventional theory, is normally quoted as

$$\nabla x E = -\frac{\partial B}{\partial t}$$

The conventional result, $W_\perp = \mu B$, can then be obtained from this by applying Stoke's theorem and integration. However, this differential form gives the electric field only for a **fixed** region in the frame in which B and E are measured and does not include the extra "convective" term. It does produve the emf due to the changing field but only in a rigid, stationary circuit. It does not give the electric field, or motional emf, in a moving circuit, which is needed in the case of a changing Larmor orbit. In this situation the differential equation requires an **extra** "convective" term which uses the relationship between the electric field in the moving circuit, **E'** and the electric field in the fixed reference frame, **E**:

$$\boldsymbol{E'} = \boldsymbol{E} + \boldsymbol{v} \: x \: \boldsymbol{B}$$

A number of authors (eg Jackson, Classical Electrodynamics) make this point about the limitation of the differential form. The rate of change of flux when the circuit is moving at velocity v, is given by Jackson and other authors as

$$\frac{\partial}{\partial t}\int_S \boldsymbol{B}.\boldsymbol{n}\:da = \int_S \frac{\partial \boldsymbol{B}}{dt}.\boldsymbol{n}\:da + \oint_C (\boldsymbol{B}\:x\:\boldsymbol{v}).d\boldsymbol{l}$$

If this formulation is used, then the same result as in equation (15) is obtained.

## 4. Implications of the new theory

**Deceleration of plasmas moving into lower field regions**

Figure 10 shows a charged particle moving in the positive direction of the z-axis, into a magnetic field that is decreasing. This is indicated by the magnetic field lines, which are diverging. Conventional theory requires that particles moving into lower field regions accelerate along the field lines, gaining in energy $W_\parallel$, but losing equally in energy perpendicular to the field, $W_\perp$. In conventional theory the velocity $V_\parallel$ along the z-axis should increase, as in the behaviour of a magnetic mirror, required by conventional theory – which the new theory suggests is incorrect.

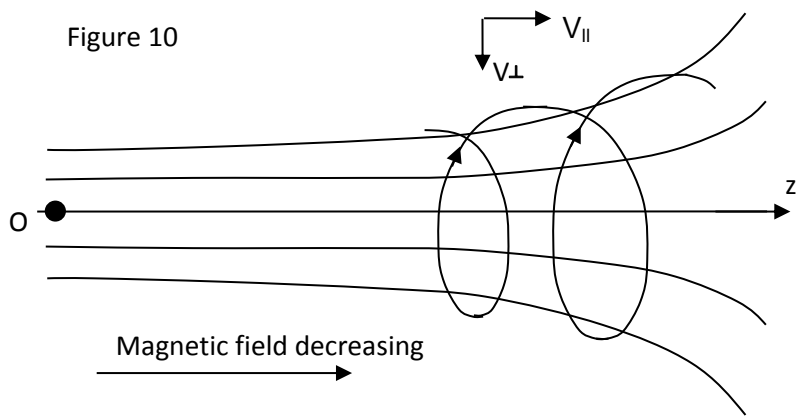

Figure 10

Magnetic field decreasing

However the new theory indicates that in this figure, the horizontal velocity $V_∥$ is **decreasing**, while the perpendicular velocity $V_⊥$ is increasing. This is exactly the opposite to conventional theory. This particle will decelerate, and if $V_∥$ is eventually reduced to zero, it will reverse its direction and accelerate back into the higher field regions.

This deceleration/acceleration process is similar to that seen in solar spicules.

**Acceleration/Deceleration equation**

The acceleration/deceleration of a charged particle can be found as follows:

The total energy of the particle is constant: $W_⊥ + W_∥ = W_{total}$

$$\text{So } \frac{dW_⊥}{dz} = -\frac{dW_∥}{dz}, \quad \text{where z is the distance along the horizontal.} \tag{17}$$

Since
$$\frac{dW_∥}{dz} = m\frac{dV_∥}{dt}$$

we get
$$m\frac{dV_∥}{dt} = \frac{^1/_2 k}{B^{3/2}}\frac{dB}{dZ} \tag{18}$$

using equation (15).

The acceleration $dV_∥/dt$ will clearly be constant if $\frac{1}{B^{3/2}}\frac{dB}{dZ}$ is constant.

(It is shown later that this term is approximately constant in particular circumstances)

k is found from equation (15) using the initial conditions, $k = W_{⊥0} B_0^{1/2}$

The initial perpendicular energy, $W_{⊥0}$ can be expressed in terms of $W_{∥0}$ if we assume for simplicity that there is an equipartition of energy at the origin, where $B = B_0$. Then we have for the average particle, simply

$$W_{⊥0} = W_{∥0}.$$

Using $= \frac{1}{2}mV^2$, equation (18) can then be rewritten as

$$\frac{dV_∥}{dt} = \frac{V_{∥0}^2 B_0^{1/2}}{4 B^{3/2}}\frac{dB}{dZ} \tag{19}$$

$V_{∥0}$ is the velocity of the plasma measured at Z=0, which is chosen as the origin, and the point at which $B=B_0$.

The mass m of the particle cancels, indicating this formula does not depend on the species of particle. Note that if dB/dz is negative, so is $dV_∥/dt$, indicating a deceleration. If we let $dV_∥/dt$ be denoted by 'a', this equation can be written:

$$a = \frac{B_0^{1/2}}{4 B^{3/2}}\frac{dB}{dZ} V_{∥0}^2 \tag{20}$$

# 5. Evaluation of the term $\frac{B_0^{1/2}}{4B^{3/2}}\frac{dB}{dZ}$

This will clearly depend on the form of equation used to describe B(z), the magnetic field. Two situations are examined, which have broadly similar results – an inverse square law and an exponential form.

**Inverse square law**

Let the magnetic field be represented, over a limited region by

$$B(z) = \frac{B_0}{(1+z/z_0)^2} \tag{21}$$

The origin, z = 0 is taken as the point at which the initial velocity, $V_{\parallel 0}$ is considered.

The **significance of $z_0$** is that it is the distance of the "origin" of the magnetic field from z=0. This can be seen by making z= - $z_0$. B becomes infinite, indicating that the source of the magnetic field is at a distance $z_0$ below the origin. For localised fields, created just around or below the photosphere, this may be as low as 5,000 km , Brosius et al (2002), but for plasmas some distance from the solar surface, it is more appropriate to use the inverse square law, with $z_0$ = Rs, the radius of the sun, Mann et al (1999).

By substituting for $B^{3/2}$ in equation (20) and finding dB/dZ from equation (21), the acceleration equation becomes, perhaps surprisingly,

$$a = -\frac{1}{2Z_0}V_{\parallel 0}^2 \tag{22}$$

**Maximum plasma height**

If the plasma reaches a maximum height, its energy parallel to z, $W_{\parallel 0,}$ will be zero and all the particle's energy will be in the plane perpendicular to z.

Assuming, as before that initially, at the base of the spicule that $W_{\perp o} = W_{\parallel 0}$

Since $W_\perp = \frac{k}{B^{\frac{1}{2}}}$ then at the maximum height $W_\perp = 2W_{\parallel 0}$

Or $$2W_{\parallel 0} = \frac{W_{\parallel 0}(1+Z_{max}/z_0)}{B_0^{\frac{1}{2}}} B_0^{1/2}$$

Which simplifies to

$$Z_{max} = Z_0 \quad \text{which is a constant.} \tag{23}$$

If a square law applies, then the height reached by the plasma is constant at a particular point, provided the magnetic field configuration does not change. The other condition is that the plasma is collisionless.

**Time of flight**

From equations (22) and (23) we can find the time taken to reach maximum height.

Using the standard equation t =v/a gives

$$t_{max} = \frac{V_{\parallel 0}}{a} = \frac{(2aZ_0)^{1/2}}{a}$$

So $$t_{max} = \left(\frac{2z_0}{a}\right)^{1/2} \tag{24}$$

This means that the time of flight is inversely related to the acceleration: plasmas which accelerate fast last for shorter periods of time.

The advantage of applying the inverse square law is the relatively simple results it produces, making it easier to see some of the basic physics involved.

**Exponential law**

For the region immediately above the photosphere, an exponential formula is more appropriate.

The magnetic field above the photosphere can be formulated by

$$B = B_0 e^{-z/z_0} \tag{25}$$

Where $B_0$ is the field at the base of the spicule, taken as the height z = 0 km.

$Z_0$ now represents the magnetic scale height of the field.

**Deceleration – Maximum velocity relation**

$$\frac{dV_\parallel}{dt} = \frac{V_{\parallel 0}{}^2 B_0{}^{1/2}}{4B^{3/2}} \frac{dB}{dZ}$$

Which becomes

$$\frac{dV_\parallel}{dt} = -\frac{e^{z/2z_0}}{4z_0} V_{\parallel 0}{}^2$$

Let acceleration = a, then if $z < 2z_0$

$$a \approx \frac{-(1 + z/2z_0)}{4z_0} V_{\parallel 0}{}^2 \tag{26}$$

The acceleration is approximately proportional to the initial velocity provided z is not too large. This is in contrast to the case for the inverse square law, where the acceleration is constant and independent of z. The non-linear term means that the acceleration will **increase** as z increases.

**Maximum plasma height**

With the same technique used as for the inverse square law,

And using $\quad W_\perp = \dfrac{k}{B^{\frac{1}{2}}}$, then at the maximum height

$$2W_{\parallel 0} = \frac{W_{\parallel 0}}{B_0^{\frac{1}{2}} e^{-z/2z_0}} B_0{}^{1/2}$$

Solving, gives the maximum height as $\quad z_{max} = 1.39\, z_0 \tag{27}$

So the height of the spicule again depends only on the scale length of the magnetic field. This could be used to provide information about the variation of the field: when the scale length is high, so will be the spicule length.

**Time of flight**

Using the same approach as for the inverse square law now gives

$$t_{max} = \frac{2z_0{}^{1/2}}{a^{1/2} e^{z/4z_0}} \tag{28}$$

Again it can be seen that provided z is not too large, there is an inverse relationship between deceleration and time of flight.

## 6. Application of the acceleration equation to solar jets and spicules.

Equation (20) can be applied to the phenomenon of solar spicules, fibrils and jets.

It is useful, initially, to use an inverse square law for the magnetic field. The actual field is more complex than this, but over part of the photosphere and corona, it can provide a reasonable approximation to the real magnetic field. The benefit of using an inverse square law is that the acceleration relationship reduces to the relatively simple equation (22). The deceleration is then a constant, given only by magnetic field parameters and the initial velocity.

**Estimating the magnetic scale length, $z_0$**

Brosius et al (2002) give one estimate of this as 0.5 x $10^9$ cm (5 x $10^3$ km). However it is well known that the solar magnetic field is quite variable and stronger fields are often quite localised. In an empirical model of the field above an active region, Gary (2001) uses dipoles with a depth as low as 500 km. For this reason, we will examine the result of two scale heights, one of 5 x $10^3$ km and a more localised one, 1 x $10^3$ km, which is still twice the thickness of the photosphere.

The acceleration equation for each of these regions then becomes, respectively

$$a = -\frac{V_{\parallel 0}^2}{2 x 5 x 10^3} \quad (29a)$$

$$a = -\frac{V_{\parallel 0}^2}{2 x 1 x 10^3} \quad (29b)$$

Figure 11 shows these two curves. The top curve is for the higher scale height (lower acceleration). The vertical dotted line is the value of solar acceleration at the surface.

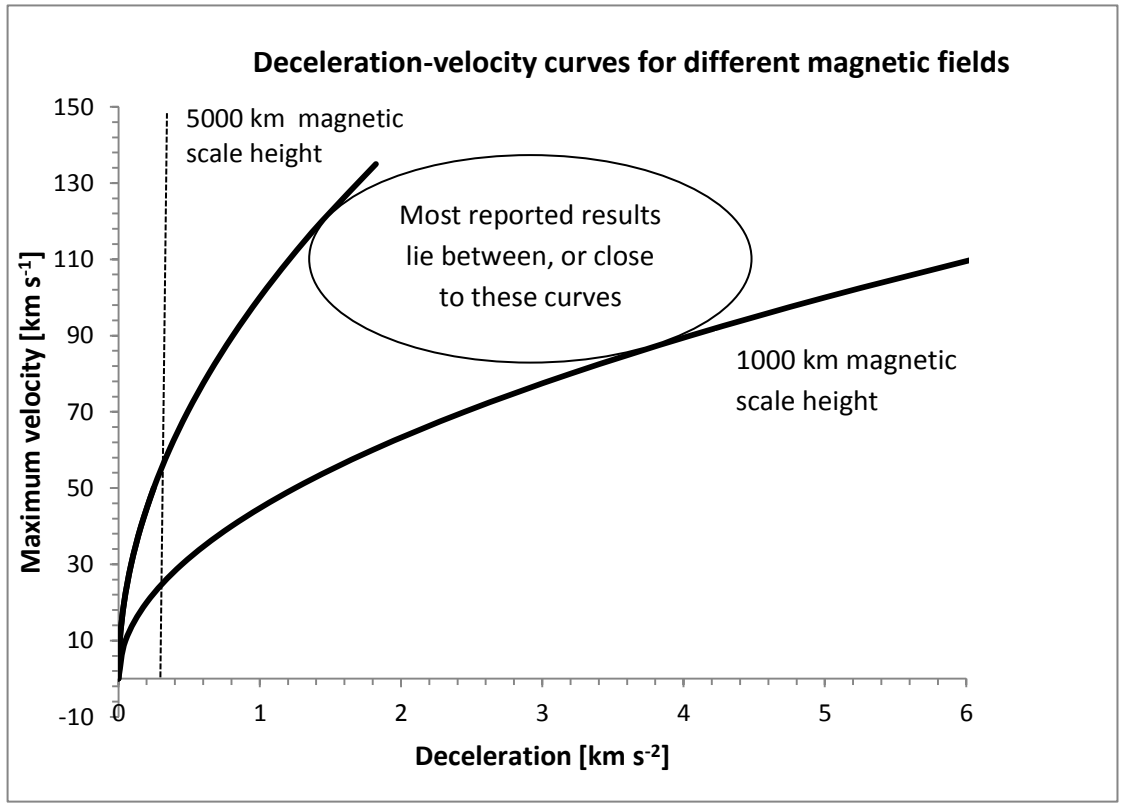

**Figure 11**

Deceleration-velocity curves for different magnetic fields

**Spicule deceleration – experimental results**

There is now substantial evidence, thanks to greatly improved spatial and temporal resolution made possible with improved instrumentation, such as Hinode/SOT and the Swedish 1m Solar Telescope, that many plasma jets, or spicules behave ballistically after ejection from the solar surface. They have a wide range of initial velocities and usually experience a constant deceleration. The deceleration is often many times greater than that due to gravity and so cannot be explained simply by solar gravity. (Solar deceleration is shown by the vertical dotted line in figure 11). At other times the deceleration is small, much less than solar gravity.
The deceleration has been found to be correlated with the maximum (that is the initial) velocity with which the plasma jet is launched from the solar surface.

This phenomenon has been described by a number of researchers: Hansteen et al (2006), De Pontieu (2007), Langangen et al (2008) Anan et al, (2009), Zhang et al (2012) and Pereira et al (2012). They made measurements of hundreds of spicular features, recording their deceleration, initial velocity, length and duration time.
Each produced scattergrams of deceleration and velocity, and showed a clear correlation between the deceleration of a spicule and its initial velocity.
Anan et al recorded decelerations up to 2.3 km s$^{-2}$ for specular jets over a plage area, close to the limb. Zhang et al had a larger range of velocities and decelerations, up to 5.5 km s$^{-2}$.
The great majority of the points in the scattergrams fitted into the area shown in figure11, between the two curves, or close to them. There were, however a number points that did not fit, showing high velocity and low acceleration, but these were in a minority

Zhang et al (2012), examined hundreds of spicules in both quiet sun and coronal hole regions. Table 1 gives their results for velocity and deceleration and shows the quantity a/V$^2$.

**Table 1**

|  | Quiet sun | Coronal hole |
|---|---|---|
| Number of spicules | 105 | 102 |
| Mean acceleration, a, (km s$^{-2}$) | -0.14 | -1.04 |
| Mean vertical velocity, V, (km s$^{-1}$) | 15.5 | 40.5 |
| **a/V$^2$ (km$^{-1}$)** | **5.8x 10$^{-4}$** | **6.3x10$^{-4}$** |

This quantity a/V$^2$ has been calculated from Zhang's results. Although there is almost a factor of 10 difference between the accelerations of quiet sun and coronal hole situations, this "acceleration constant" is essentially the same in both cases. The mean value is approximately 6.0 x 10$^{-4}$ km$^{-1}$
This will produce a curve just under the lower curve in figure 11 and is a reasonable fit to much of their data.

**Maximum height**

From equation (23) and using the empirical value in equation (29b) gives z$_{max}$ = 1000 km, which is a typical value for the spicule length found by Anan (2010), so the new theory does provide some explanation for the length of spicules. The length of a spicule reflects the local magnetic scale length over the solar surface: a field which is strongly localised will produce short, fast spicules.

## Time of flight: support for a = kV$^2$

The time of flight of spicules is experimentally inversely related to deceleration, according to Pereira et al (2012). If the acceleration were to directly proportional to the initial velocity, then it is easy to show that the time of flight would be a constant. However, the inverse relationship shown by the scattergram of Pereira et al does closely follow the inverse square root relation seen in equation (24) of the new theory.

Figure 12 shows a graph of this equation, where the constant $Z_0$ has been chosen as 5,000 km. The shaded area approximates to the scattergram of results for the deceleration-time graph of Pereira et al, shown in their figure 7. The maximum height $Z_0$ is then 5,000 km, which is in reasonable agreement with the maximum heights they found.

Figure 12

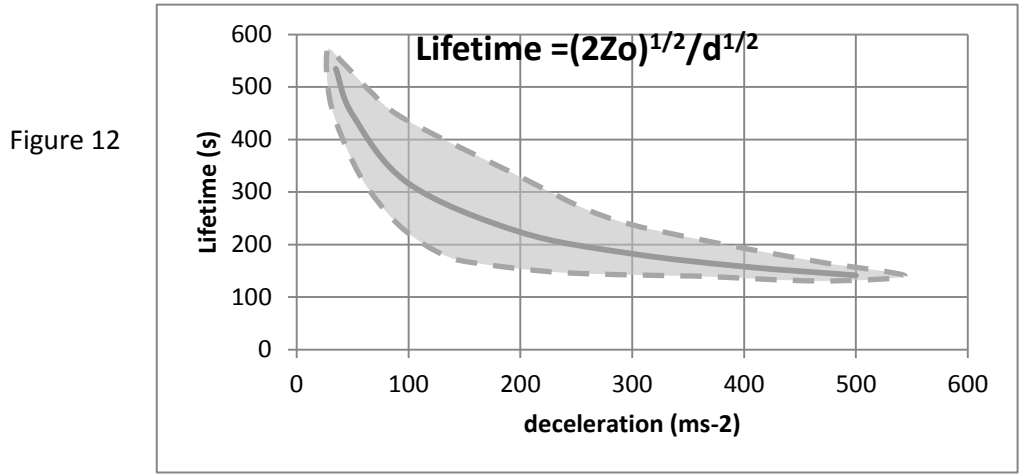

Lifetime $=(2Z_0)^{1/2}/d^{1/2}$

## 8. Variation of magnetic moment

Conventional theory states that the magnetic moment of a charged particle is constant, provided the magnetic field changes slowly. But there is some evidence that the magnetic moment of a plasma such as the solar wind increases as the magnetic field decreases, which may support the new theory. In www.livingreviews.org/lrsp-2006-1, for example Marsch reports that the magnetic moment of protons increases by a factor of about 3 in moving out from 0.4 $R_0$ to 0.98 $R_0$, approximately. This may be partly due to the increase in magnetic moment caused by decreasing fields, as seen in equation (16), which predicts a factor of 3.8 increase for these distances.

## 9. Type I and type II solar spicules

There has been research into the existence of two types of solar spicules recently (Pereira et al (2012)), Anan et al (2012)) and of the role of type II spicules in heating the solar corona. (De Pontieu et al (2007) and Klimchuk (2012)). Type II are generally shorter lived and more energetic. If the new theory presented here is correct, this may simply be related to the "magnetic height", $Z_0$ of the magnetic field which is constraining the spicules. Type II may be those where the plasma is in a very localised magnetic field, with lower values of $Z_0$ and type I correspond to spicules that are in a more extended field, with larger $Z_0$.

## 10. CME ( Mass Ejection) and CBF (Coronal Bright Front) deceleration

It is well established that after its initial propulsion from the solar surface, a CME will generally decelerate between the sun and earth. If the new theory is correct, then some of this deceleration will be due to the deceleration process here. The process is complicated by the presence of the solar wind, however, which will change the velocity of the CME. When a CME completes its deceleration, its velocity is reduced to that of the solar wind.

Reiner et al (2007) analysed 42 CMEs and found a reasonably constant deceleration, correlated to the initial velocity of the CME. The power-law fit they found between deceleration and the initial CME/shock speed was : $-a(\text{m s}^{-2}) = 4.55 \times 10^{-8} v_0(\text{km s}^{-1})^{2.61}$.

They also found that the initial CME/shock speed was approximately inversely proportional to the square root of the duration time, $V_{\parallel 0} \propto t^{-0.477}$, as required by the theory proposed here.

Gallagher et al (2011) analysed the kinematics of CBFs (Coronal Bright Fronts), associated with CMEs. They comment on a CBF of 19 May, 2007 which after its initial acceleration stage, had "a start velocity of 460 km s$^{-1}$ with a constant deceleration of $-160$ ms$^{-2}$".
Using $L = V^2/2a$, gives a "magnetic length" of $0.65 \times 10^6$ km. This appears to happen, from their figure 7, at around a height of $0.4 \times 10^6$ km. As these two lengths are of the same order of magnitude, magnetic deceleration may play a significant factor in decelerating CBFs.

## 11. Discussion and summary

The new theory of magnetic moment predicts behaviour which is almost opposite to the old theory. Despite this, it does seem to produce a reasonable explanation for the velocity- deceleration relationship, the length of solar spicules and the deceleration-time relationship despite a number of simplifying assumptions. One of these assumptions is that the plasma is largely collisionless. If there are collisions during acceleration/deceleration, this will tend to make the particle's energy more isotropic. For higher density spicules, this will mean higher velocities, $V_\parallel$, and the spicule will then rise to greater heights. This could be one of the reasons for the minority of experimental measurements that lie well above the curves in figure 11. Statistical data on spicule accelerations and velocities is limited in its usefulness, since the value of $Z_0$ will vary from spicule to spicule. Some way of comparing $W_\perp$ and $W_\parallel$ along individual spicules could be very valuable in assessing the validity of the theory proposed, as could the variation of magnetic field along a spicule.
Different forms for the magnetic field have been used. Although the detailed results they give vary, the main characteristics are broadly the same, irrespective of the type of magnetic field formula.
CMEs also appear to experience the same type of deceleration, and it is likely that this is partly due to the magnetic deceleration proposed by this new theory.
Solar gravity has been ignored in this paper. This may be partly justified as solar gravitational acceleration is much less than the values shown in figure 11, and the fact that many of the spicules are not vertical. The effect of solar acceleration will be to add an extra term to the overall deceleration, depending on the inclination of the spicule.
The other assumption to be considered is that, at the point at which the initial velocity is measured, the energy of the particle is equipartitioned. If the particle has been travelling, collision-free for some time in the magnetic field before the point at which the start velocity is measured, then this assumption must be modified.

Leonard Freeman, M Inst P
23 Hope Street
Cambridge
CB1 3NA
UK